\documentclass[twocolumn,preprintnumbers,nsr,superscriptaddress]{revtex4}%
\usepackage{amsfonts}
\usepackage{amsmath}
\usepackage{amssymb}
\usepackage{graphicx}%
\begin{document}
\title{Topological quantum states of matter in iron-based superconductors: From
concepts to material realization}
\author{Ning Hao}
\email{haon@hmfl.ac.cn}
\affiliation{Anhui Province Key Laboratory of Condensed Matter Physics at Extreme
Conditions, High Magnetic Field Laboratory of Chinese Academy of Sciences,
Hefei 230031 China}
\author{Jiangping Hu}
\email{jphu@iphy.ac.cn}
\affiliation{Beijing National Laboratory for Condensed Matter Physics, and Institute of
Physics, Chinese Academy of Sciences, Beijing 100190, China}
\affiliation{CAS Center of Excellence in Topological Quantum Computation and Kavli
Institute of Theoretical Sciences, University of Chinese Academy of Sciences,
Beijing, 100190, China}
\affiliation{Collaborative Innovation Center of Quantum Matter, Beijing, 100871, China}
\date{\today}

\begin{abstract}
We review recent progress in the explorations of topological quantum states of
matter in iron-based superconductors. In particular, we focus on the
nontrivial topology existing in the band structures and superconducting states
of iron's 3d orbitals. The basic concepts, models, materials and experimental
results are reviewed. The natural integration between topology and
high-temperature superconductivity in iron-based superconductors provides
great opportunities to study topological superconductivity and Majorana modes
at high temperature.


\end{abstract}
\maketitle

%\affiliation{Collaborative Innovation Center
%of Advanced Microstructures, Nanjing University, Nanjing 210093 China}

\section{Introduction}

In the past decade, topology becomes an essential ingredient to classify
various types of materials, including insulators/semiconductors, semimetals
and superconductors\cite{Hasan_rmp_2010,Qi_rmp_2011,Armitage_rmp_2018}. The
physical consequence in a topological material is the existence of
topologically protected surface states, which can be measured directly in
transport, angle resolved photoemission spectrum(ARPES), scanning tunneling
microscopy(STM) and other
experiments\cite{Hasan_rmp_2010,Qi_rmp_2011,Armitage_rmp_2018}. In particular,
in a topological superconductor, there are surface bound states, Majorana
modes, which can be used to realize topological quantum computing because of
their topological protection and non-Abelian braiding
statistics\cite{Nayak_rmp_2008}.

While naturally-born topological superconductors are very rare, the
realization of Majorana modes can be achieved in many artificial hybrid
systems. Recently, a wealth of proposals for such experimental designs has
been proposed, including the superconducting surface states of a topological
insulator in proximity to conventional superconductors\cite{Fu_prl_2008},
quantum wires with strong spin-orbit coupling in proximity to conventional
superconductors\cite{Lutchyn_prl_2010}, semiconductor-superconductor
heterostructures\cite{Sau_prl_2010}, and spin-chains embedded in conventional
superconductors\cite{Stevan_science_2014} etc. However, these hybrid systems,
in general, have two shortcomings. First, it is always difficult to manage the
interface between two different structures. Second, in all these proposals, as
the proximity effect requires a long superconducting coherent length, high
temperature superconductors, such as cuprates and iron-based superconductors,
have never been candidates in those integration processes because of their
extreme short coherent lengths and structural incompatibility. Thus, all
devices require to be operated at very low temperature.

The above shortcomings can be overcome if we can find a high temperature
superconductor which hosts nontrivial topological band structures.
Specifically, to differentiate them from topological superconductors as well
as the above hybrid superconducting systems, we refer this type of
superconductors specifically as \textit{connate} topological
superconductors\cite{Shi_sb_2017}. The connate topological superconductor can
be viewed as an internal hybrid system which has conventional
superconductivity in bulk but topological superconductivity on surface caused
by the nontrivial topology on some part of band
structures\cite{Shi_sb_2017,Xu_prl_2016}. Because of this intrinsic
hybridization, the superconductor, in general, must be a multiple band
electronic system. As iron-based high temperature superconductors are known to
be multi-orbital electronic systems, they become promising candidates.

During the past several years, starting from theoretical understanding, the
research of iron-based superconductors as connate topological superconductors
has gradually been materialized. The first theoretical study of nontrivial
band topology was carried out by us for the single layer FeSe/STO, in which a
band inversion can take place at M points\cite{Hao_prx_2014} to create
nontrivial topology. Very quickly, it was found that the band inversion can
easily take place at $\Gamma$ point if the anion height from Fe layers are
high enough. For FeSe, the height can be increased by substituting Se with
Te\cite{Wu_prb_2016,Wang_prb_2015}. For iron-pnicitides, the As height is
predicted to be high enough in the 111 series, LiFeAs to host nontrivial
topology\cite{Zhang_arxiv_2018-1}. Besides these intrinsic topological
properties from the Fe d-orbitals, nontrivial topology can also stem from
bands outside Fe layers. For example, the As p-orbitals in the As layers of
the 122 CaFeAs$_{2}$ are shown to be described by a model similar to the
Kane-Mele model in graphene\cite{Wu_prb_2015-1}. Most recently, because of the
improvement of sample quality and experimental resolutions, there have been
increasing experimental evidence for topological properties in iron-based
superconductors\cite{Zhang_science_2018,Wang_science_2018,Liu_arxiv_2018}. The
theoretically predicted band inversions, together with the topologically
protected surface states, have been directly observed. The Majorana-like modes
are observed in several iron-chalcogenide
materials\cite{Wang_science_2018,Liu_arxiv_2018}. All these progresses have
made iron-based superconductors to be a new research frontier for topological superconductivity.

In this paper, we give a brief review of both theoretical and experimental
results regarding of the topological properties of iron-based superconductors.
In section II, we discuss theoretical concepts and models for the topological
band structure in iron-based superconductors and recent experimental evidence.
In section III, we review topological superconductivity that can be emerged
from the topological bands of iron-based superconductors and experimental
evidence of Majorana-like modes in these materials. Finally, we will address
open issues in this field.

\section{Topology in iron d-orbital bands}

\subsection{Concepts and models}

\begin{figure*}[ptb]
\begin{center}
\includegraphics[width=1.0\linewidth]{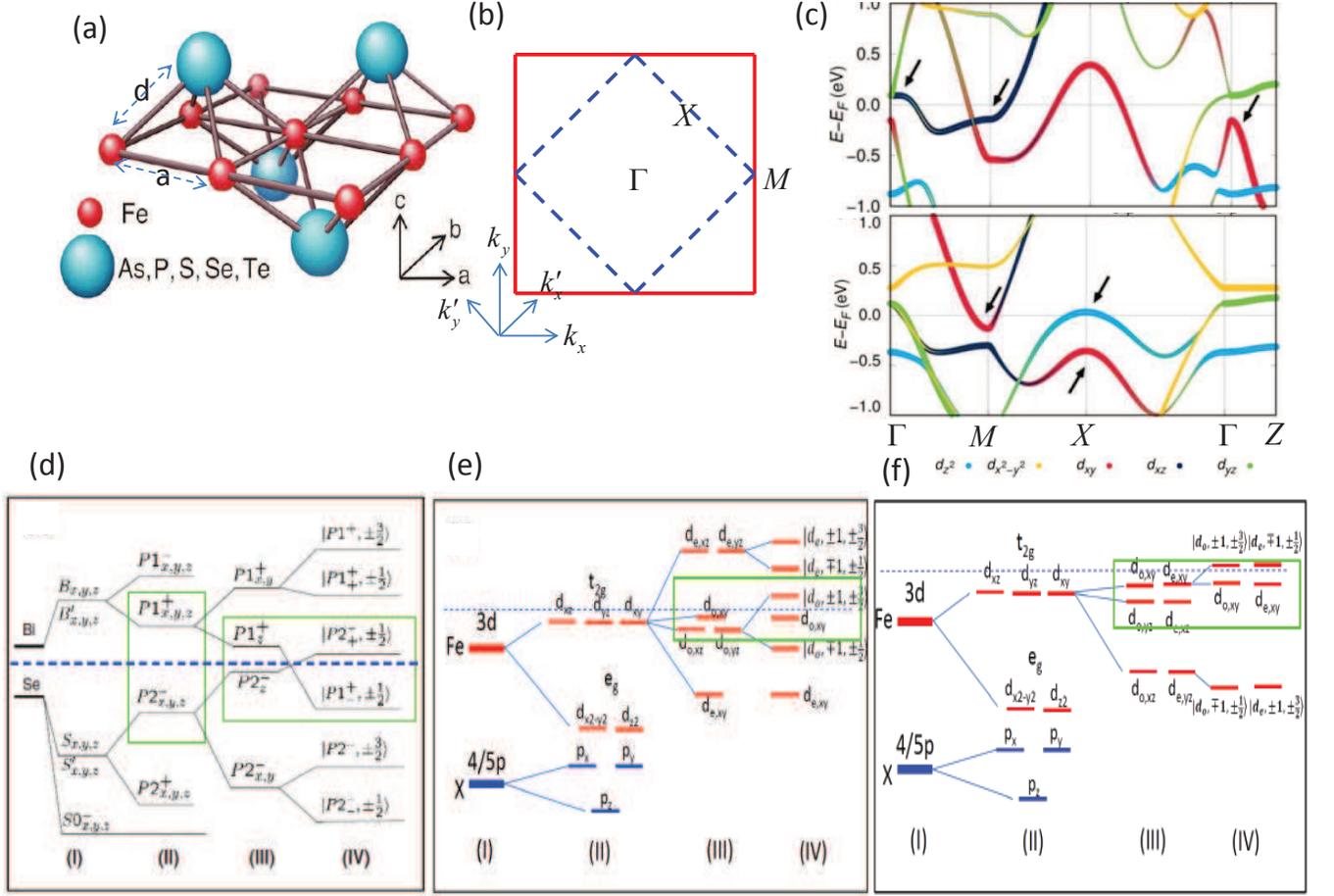}
\end{center}
\caption{(Color Online) (a) The structure of X-Fe-X trilayer. The distance
between the nearest neighbor two ions/ion and X is label with $a$ and $d$.
(Adopted from \cite{Podolsky-1}) (b) The Brillouin zone with high-symmetry
point. The red solid/blue dashed lines label the Brillouin zone with
one-iron/two-iron unit cell. In (c) the top/bottom panels correspond to
parameters ($a$, $d$)=(0.93,0.98)/(1.09,1) in unit of experimental values of
FeSe\cite{Guterding_prb_2017}. (d) Schematic picture of the origin of the band
structure of Bi$_{2}$Se$_{3}$. Starting from the atomic orbitals of Bi and Se,
the following four steps are required to understand the band structure: (I)
the hybridization of Bi orbitals and Se orbitals, (II) the formation of the
bonding and antibonding states due to the inversion symmetry, (III) the
crystal field splitting, and (IV) the influence of the spin-orbit coupling,
from \cite{Liu_prb_2010}. (e) and (f) the similar processes in iron-based
superconductors at high-symmetry point $\Gamma$ in (e) and point $M$ in (f),
from \cite{Hao-2018}. In both (e) and (f), (I) the hybridization of iron 3d
orbitals and X 4p or 5p orbitals, (II) the crystal field splitting, (III) the
formation of the bonding and antibonding states, which are classified with the
parities of glide-plane symmetry, and (IV) the influence of the spin-orbit
coupling or other effects.}%
\label{fig1}%
\end{figure*}

Since the discovery of iron-based superconductors in 2008, there has been
remarkable progress in material growth and synthesis about the iron-based
compounds. According to the element composition, the iron-based
superconductors are classified into different categories denoted with
\textquotedblleft1111\textquotedblright, \textquotedblleft%
122\textquotedblright, \textquotedblleft111\textquotedblright,
\textquotedblleft11\textquotedblright, etc\cite{Paglione_np_2010}. All
categories possess the kernel substructure of X-Fe-X trilayer with X denoting
As, P, S, Se, Te, as shown in Fig.\ref{fig1} (a). The X-Fe-X trilayer is the
basic unit cell to give arise to magnetism and superconductivity, and play a
similar role as Cu-O plane in cuprates. Following the principle from
complexity to simplicity, the X-Fe-X trilayer skips the specificity among all
the compounds in iron-based superconductors and brings the intrinsic physics
to the surface. However, along the opposite logic, the diversity may include
important subtle surprising differences. For iron-based superconductors, such
kinds of accidental surprises can be intuitively demonstrated through
evaluating the sensitivity of the electronic structures upon the tiny change
of the structure of the X-Fe-X trilayer\cite{Guterding_prb_2017}.
Fig.\ref{fig1} (c) gives such intuitive demonstration. The band structures
sensitively depend on the fine tune of the distances between Fe-Fe and Fe-X.
In particular, the bands switch orders near $\Gamma$ point, a band gap opens
near $M$ point and the bands become strongly dispersive along $\Gamma-Z$
direction when the third dimension is considered. Indeed, the layered
structures of the iron-based superconductors provide the possibilities to tune
the distances between Fe-Fe and Fe-X. For example, the La-O layer in LaOFeAs
and the Ba-As layer in BaFe$_{2}$As$_{2}$ naturally cause different lattice
constants for Fe-X layers\cite{Singh_prl_2008,Singh_prb_2008}. A variety of
materials in the family of iron-based superconductors provide different
fine-tuned X-Fe-X trilayers. 

The band fine tuning would become nontrivial if there exist a topological
phase transition. The discovery of topological insulators has established a
standard paradigm about the topological quantum states of matter, which
includes band inversion, bulk-boundary correspondence and relationship between
symmetry and topological invariant
etc\cite{Kane_prl_2005-1,Bernevig_science_2006,Konig_science_2007,Moore_prb_2007,Fu_prl_2007,Roy_prb_2009,Fu_prb_2007,Chen_science_2009,Hasan_rmp_2010,Qi_rmp_2011}%
. For example, the first experimentally confirmed two-dimensional topological
insulator, HgTe/CdTe quantum well, has a band inversion induced by the large
spin-orbit coupling from Hg, depending on the thickness of the well, to gives
arise to a topological insulating
state\cite{Bernevig_science_2006,Konig_science_2007}. The well-known
three-dimensional topological insulators, Bi$_{2}$Se$_{3}$ and Bi$_{2}$%
Te$_{3}$, has a band inversion caused by a strong spin-orbit coupling that
switch two p$_{z}$-type bands with opposite spacial-inversion-symmetry
parities at the $\Gamma$
point\cite{Zhang_np_2009,Chen_science_2009,Xia_np_2009}. The picture of band
inversion can be further simplified into the energy level shift in atomic
limit through the adiabatic deformations\cite{Zhang_np_2009}. Fig. \ref{fig1}
(d) gives the typical picture of energy level shift under the influence of
several kinds of interactions in Bi$_{2}$Se$_{3}$\cite{Liu_prb_2010}.

Interestingly, a similar picture also exists in some specific iron-based
superconductors with fine-tuned X-Fe-X layers. The typical picture of energy
level shift of iron $d$-orbitals are shown in Fig. \ref{fig1} (e) and (f) for
$\Gamma$ point in (e) and $M$ point in (f), respectively. Note that the space
group of Fe-X-Fe trilayer is P4/nmm, in which the glide-plane mirror symmetry
operation \{$m_{z}|\frac{1}{2}\frac{1}{2}0$\} and inversion symmetry operation
\{$i|\frac{1}{2}\frac{1}{2}0$\} are
essential\cite{Cvetkovic_prb_2013,Hao_prb_2014,Hao_prx_2014,Hao_prb_2015}.
First, the Bloch states can be classified according to the parities of
\{$m_{z}|\frac{1}{2}\frac{1}{2}0$\}, i.e., $|d_{o/e,\alpha}\rangle$ or
$|d_{o/e},m_{l},m_{j}\rangle$ with $o$, $e$, $\alpha$, $m_{l}$, $m_{j}$
denoting the odd or even parity of \{$m_{z}|\frac{1}{2}\frac{1}{2}0$\}, the
$\alpha$th $d$ orbital, and two magnetic quantum numbers, respectively.
Second, under the inversion symmetry operation \{$i|\frac{1}{2}\frac{1}{2}%
0$\}, the inversion parities of $|d_{o/e,\alpha}\rangle$ and $|d_{o/e}%
,m_{l},m_{j}\rangle$ for $\alpha=xz/yz$, $m_{l}=\pm1$ are opposite to the
inversion parities of $|d_{o/e,\alpha}\rangle$ for $\alpha=xy$. Focus on the
green rectangles in Fig. \ref{fig1} (e) and (f), the spin-orbit coupling can
switch the order of the energy levels with opposite inversion parities and
induce a topological phase
transition\cite{Hao_prb_2014,Hao_prx_2014,Hao_prb_2015}.

In early 2014, the authors of this paper noted that a tiny band gap around $M$
point in the band structure of monolayer FeSe/SrTiO$_{3}$%
(FeSe/STO)\cite{Wang_cpl_2012} from measurement of the
ARPES\cite{He_nm_2013,Tan_nm_2013,Liu_nc_2012,Lee_nature_2014,Miyata_nm_2015,Peng_nc_2014}
and predicted the topological phase transition in this two-dimensional system
\cite{Hao_prx_2014}. It is the first proposal to discuss the topological
quantum state of matter in iron-based superconductors. Corresponding to Fig.
\ref{fig1} (f), an effective \textbf{k}$\cdot$\textbf{p} model can be
constructed in the basis set of $[\{|\psi_{o}\rangle\},\{|\psi_{e}\rangle\}]$
with $\{|\psi_{o}\rangle\}=\{|d_{o,xy,\uparrow}\rangle,|d_{o},1,\frac{3}%
{2}\rangle,|d_{o,xy,\downarrow}\rangle,|d_{o},-1,-\frac{3}{2}\rangle\}$ and
$\{|\psi_{o}\rangle\}=\{|d_{e,xy,\uparrow}\rangle,|d_{e},-1,\frac{1}{2}%
\rangle,|d_{e,xy,\downarrow}\rangle,|d_{e},1,-\frac{1}{2}\rangle\}$%

\begin{equation}
H_{M}(k)=\left[
\begin{array}
[c]{cc}%
H_{M}^{o}(k) & H_{c}\\
H_{c} & H_{M}^{e}(k)
\end{array}
\right]  . \label{HM1}%
\end{equation}
Here,%

\begin{equation}
H_{M}^{o}(k)=\left[
\begin{array}
[c]{cc}%
H(k) & 0\\
0 & H^{\ast}(-k)
\end{array}
\right]  , \label{Hm2}%
\end{equation}
$H_{M}^{e}(k)=$ $H_{M}^{o,\ast}(-k)$, $H(k)=\varepsilon(k)+d_{i}(k)\sigma_{i}$
with $\varepsilon(k)=C-D(k_{x}^{2}+k_{y}^{2})$, $d_{1}(k)+id_{2}%
(k)=A(k_{x}+ik_{y})$, and $d_{3}(k)=M-B(k_{x}^{2}+k_{y}^{2})$ with $MB>0$. In
the absence of $H_{c}$ term, the Hamiltonian in Eq.(\ref{HM1}) reduces into
two copies of Bernevig-Hughes-Zhang (BHZ) model\cite{Bernevig_science_2006},
which is the standard model for quantum spin Hall effect. In each subspace
with odd or even parity, a topological invariant $Z_{2}=1$ can be defined.
Actually, $H_{c}$ term is from the spin-flipped term $\lambda_{so}(L_{x}%
s_{x}+L_{y}s_{y})$, which mixes the orbitals with odd and even parities of
\{$m_{z}|\frac{1}{2}\frac{1}{2}0$\}. As a consequence, the parity of
\{$m_{z}|\frac{1}{2}\frac{1}{2}0$\} is no longer a good quantum number. The
two subspaces couple with each other. The topological states is more like weak
type. However, if the two iron sublattices have different on-site potential,
i.e., the staggered sublattice potential, which is introduced by the
substrate, the weak topological state can be tuned into the strong topological
states, because the potential can renormalize the mass term $M$ in $d_{3}(k)$,
and change its sign in only one copy. Now, the band inversion condition with
$MB>0$ is satisfied only in another copy. The topological state becomes strong
type and is robust against the $H_{c}$ coupling without breaking time reversal
symmetry\cite{Hao_prx_2014}.

In late 2014, the topological phase transition around $\Gamma$ point was
proposed in Fe(Te$_{1-x}$Se$_{x}$) thin film\cite{Wu_prb_2016}, as well as in
the bulk materials\cite{Wang_prb_2015}. The first-principle calculations
indicated that the proper ratio between Te and Se could induced the band
inversion around $\Gamma$ point. Refer to Fig. \ref{fig1} (e), an effective
\textbf{k}$\cdot$\textbf{p} model can be constructed in the basis set
$\{|d_{o},1,\frac{3}{2}\rangle,|d_{o,xy,\uparrow}\rangle,|d_{o},-1,-\frac
{3}{2}\rangle,|d_{o,xy,\downarrow}\rangle\}$,%

\begin{equation}
H_{\Gamma}(k)=\varepsilon_{0}+\left[
\begin{array}
[c]{cccc}%
-M(k) & Ak_{+} &  & \\
Ak_{-} & M(k) &  & \\
&  & -M(k) & -Ak_{-}\\
&  & -Ak_{+} & M(k)
\end{array}
\right]  . \label{HG}%
\end{equation}
Here, $\varepsilon_{0}=C-D(k_{x}^{2}+k_{y}^{2})$, $M(k)=M-B(k_{x}^{2}%
+k_{y}^{2})$. In the band inversion regime, $MB>0$. Likewise, effective
\textbf{k}$\cdot$\textbf{p} model around the $\Gamma$ point in Eq. (\ref{HG})
restores the famous BHZ model which describes the quantum spin Hall effect in
HgTe/CdTe quantum well. In the original paper\cite{Wu_prb_2016}, the author
considered the hybridization between $p$ orbitals of Te/Se with $d$ orbitals
of Fe. The basis functions for the \textbf{k}$\cdot$\textbf{p} model would be
complex. Here, we use the only $d$ orbitals of Fe to construct the basis
functions through downfolding the $p$ orbital parts without changing the
symmetries. Therefore, the effective \textbf{k}$\cdot$\textbf{p} models in the
basis sets involving $d$ and $p$ orbitals or only $d$ orbitals have the
identical forms.

The topological phase transition around $\Gamma$ point in the Fe(Te$_{1-x}%
$Se$_{x}$) thin film can be generalized into the bulk Fe(Te$_{1-x}$Se$_{x}$)
single crystal. Correspondingly, the two-dimensional topological state is
generalized into three-dimensional topological states, which is similar to
topological insulator in Bi$_{2}$Se$_{3}$. The topological nature of the band
structures of bulk Fe(Te$_{1-x}$Se$_{x}$) single crystal was proposal through
the first-principles calculations\cite{Wang_prb_2015}. The band inversion and
$Z_{2}$ topological invariant was revealed. Following the picture of
topological phase transition at $\Gamma$ point shown in Fig. \ref{fig1} (e),
the topological phase transition in bulk Fe(Te$_{1-x}$Se$_{x}$) single crystal
is a little different from that in FeTe$_{1-x}$Se$_{x}$ thin film. The
spin-orbit coupling in the latter case does not play a primary role to the
topological phase transition\cite{Wu_prb_2016}. The spin-orbit coupling,
however, is indispensable in the former case. Because the small band gap
between $\Gamma_{6}^{+}$ and $\Lambda_{6}$ between $\Gamma-Z$ points is from
the \textquotedblleft transmission effect\textquotedblright, which transmits
the coupling between $\Gamma_{4}^{+}$ and $\Gamma_{5}^{+}$ to the coupling
between $\Gamma_{6}^{+}$ and $\Lambda_{6}$ through the medium of spin-orbit
coupling (See Ref.\cite{Wang_prb_2015} for the relevant band labeling). The
\textquotedblleft transmission effect\textquotedblright\ can be revealed by a
tight-binding model only involving the five $d$ orbitals of irons (the weight
of $|p_{z},\mathbf{k}+\mathbf{Q}\rangle$ state in $\Gamma_{2}^{-}$ band can be
renormalized to the $|d_{xy},\mathbf{k}\rangle$ state). The interlayer
couplings include the parity-conserved terms and the parity-mixing
terms\cite{Hao_prx_2014}. Note that the $\Gamma_{2}^{-}$ state in the
first-principles calculations is captured by the band 4 in Fig. \ref{fig2}(i).
Without interlayer parity-mixing term, even the spin-orbit coupling cannot
open a gap between band 4 and bands 1, 2. Only when both interlayer
parity-mixing term and spin-orbit coupling are tuned on, a small band gap
opens as shown in Fig. \ref{fig2}(i). The key interlayer parity-mixing term is
the hopping between the $d_{xz}$ and $d_{yz}$, i.e., $-4it_{xz,yz}^{c}(\cos
k_{x}+\cos k_{y})\sin k_{z}$. The effect of interlayer parity-mixing term can
be renormalized to obtain an effective spin-orbit coupling under the
second-order perturbation approximation,%

\begin{equation}
\tilde{H}_{soc}=\left[
\begin{array}
[c]{cc}%
0 & \tilde{h}_{soc}\\
\tilde{h}_{soc}^{\dag} & 0
\end{array}
\right]  , \label{Hsoc_eff}%
\end{equation}%
\begin{equation}
\tilde{h}_{soc}\propto\lambda_{soc}[H_{c}^{\dag}L^{-}+L^{-}H_{c}].
\label{Hsoc_eff1}%
\end{equation}
Here, $\lambda_{soc}$ is the strength of spin-orbit coupling. $L^{-}$ is the
matrix of $d$ orbitals. $H_{c}$ is the interlayer parity-mixing term. Along
the $\Gamma-Z$ line, $(k_{x},k_{y})=(0,0)$, we have%

\begin{equation}
\tilde{h}_{soc}\propto it_{xz,yz}^{c}\lambda_{soc}\sin k_{z}\left[
\begin{array}
[c]{ccccc}%
0 & 0 & -i & 1 & 0\\
0 & 0 & -1 & -i & 0\\
-i & -1 & 0 & 0 & -\sqrt{3}\\
1 & -i & 0 & 0 & \sqrt{3}i\\
0 & 0 & -\sqrt{3} & \sqrt{3}i & 0
\end{array}
\right]  . \label{Hsoc_eff2}%
\end{equation}
Based on the information from the tight-binding Hamiltonian, the effective
\textbf{k}$\cdot$\textbf{p} Hamiltonian around the $\Gamma-Z$ line can be
constructed under the basis spanned by the states $|1\rangle$, $|2\rangle$,
$|3\rangle$, and $|4\rangle$ in Fig. \ref{fig2}(i)\cite{Xu_prl_2016}. The
detailed form of the effective \textbf{k}$\cdot$\textbf{p} Hamiltonian can be
constructed in the basis set, $[\{|\psi_{\uparrow}\rangle\},\{|\psi
_{\downarrow}\rangle\}]$ with $\{|\psi_{\uparrow}\rangle\}=\{|d_{e,xy,\uparrow
}\rangle,|d_{o},1,\frac{3}{2}\rangle,|d_{o},-1,-\frac{1}{2}\rangle
,|d_{o,xy,\uparrow}\rangle,\}$ and $\{|\psi_{\downarrow}\rangle
\}=\{|d_{e,xy,\downarrow}\rangle,|d_{o},1,\frac{1}{2}\rangle,|d_{o}%
,-1,-\frac{3}{2}\rangle,|d_{o,xy,\downarrow}\rangle\},$%

\begin{align}
H_{\Gamma Z}(k)  &  =\left[
\begin{array}
[c]{cccc}%
M_{1}(k) & \gamma\sin k_{z}k_{-} & \gamma\sin k_{z}k_{+} & 0\\
& M_{2}(k) & \alpha k_{+}^{2}+\beta k_{-}^{2} & i\delta k_{-}\\
&  & M_{2}(k) & i\delta k_{+}\\
&  &  & M_{4}(k)
\end{array}
\right]  \otimes I_{2\times2}\nonumber\\
&  +H_{\Gamma Z}^{soc}(k) \label{Hgz}%
\end{align}
Here, the mass terms $M_{n}(k)=E_{n}+\frac{k_{\parallel}^{2}}{2m_{nx}}%
+t_{nz}(1-\cos k_{z})$ with $n=1$, 2, 4. $H_{\Gamma Z}^{soc}$ are some
components of $\tilde{h}_{soc}$ in Eq. (\ref{Hsoc_eff2}) and have the
following form, $H_{\Gamma Z}^{soc}(k)=[h_{11},h_{12};h_{12}^{\ast},-h11]$
with $h_{11}=\frac{\lambda_{soc}}{2}[(\sigma_{z}-1)\oplus(\sigma_{z}+1)]$,
$h_{12}=\frac{\sqrt{2}\lambda_{soc}}{2}[i\sigma_{x}-\sigma_{y},(1-\sigma
_{z})k_{z};1+\sigma_{z},(i\sigma_{x}+\sigma_{y})k_{z}]$.

Note that the band 4 and band 2 cross along $\Gamma-Z$ line without gap
opening in Fig. \ref{fig2} (i). Actually, this can be called topological Dirac
semimetal states when the chemical potential is moved to the cross point. This
state can also be described by the effective model in Eq. (\ref{Hgz}). The
target materials include Fe(Te,Se) and
Li(Fe,Co)As\cite{Zhang_arxiv_2018,Zhang_arxiv_2018-1}.

\subsection{Materials and experiments}

\begin{figure*}[ptb]
\begin{center}
\includegraphics[width=1.0\linewidth]{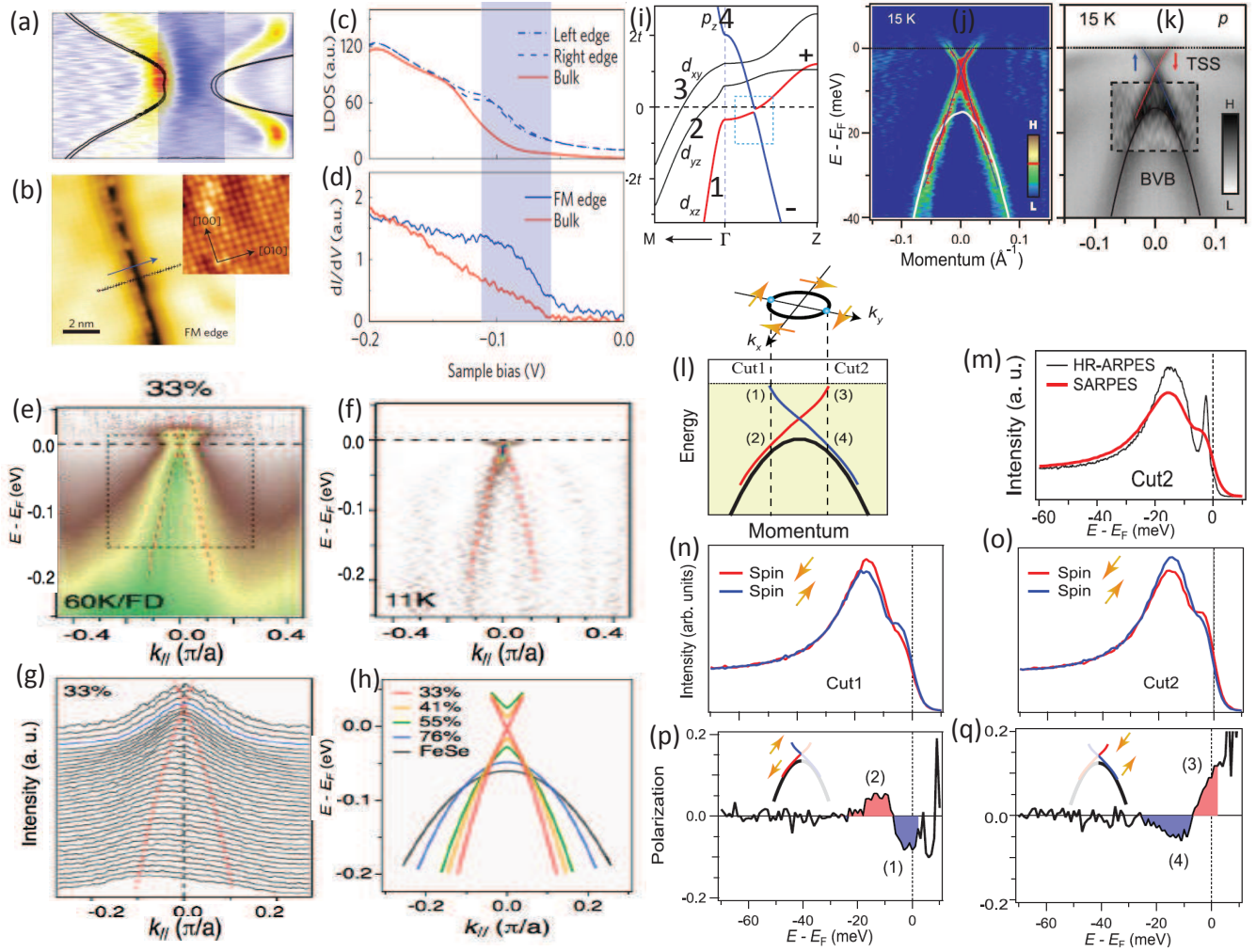}
\end{center}
\caption{(Color Online)(a)-(d) The ARPES and STM experimental results for
monolayer FeSe/STO\cite{Wang_nm_2016-1}. (e)-(f) The ARPES experimental
results for monolayer Fe(Te$_{1-x}$Se$_{x}$)/STO\cite{Shi_sb_2017}. (i)-(q)
The ARPES experimental results for bulk Fe(Te,Se) single
crystal\cite{Zhang_science_2018}. (a) ARPES band structure around the $M$
point. The black lines are theoretical band structures. (b) Experimental STM
topography of the FM edge (0.1 nA,-300 mV) of FeSe/STO. The inset shows an
atomic-resolution STM topography image at the bulk position of the FM edge
(0.1 nA,100 mV), showing the topmost Se atom arrangement (the crystal
orientations are labelled). (c) Theoretical local density of states (LDOS) for
edge and bulk states (d) Experimental STS spectra of edge and bulk states
extracted from FM edges. The light blue band in (a), (c), (d) indicates the
SOC gap. (e) The intensity plot divided by the Fermi-Dirac distribution
function near $\Gamma$ along the $\Gamma$-$M$ direction for monolayer
Fe(Te$_{1-x}$Se$_{x}$)/STO. (f) Curvature intensity plots along the same cut
as in (e). The data were recorded at the temperature indicated in the panel.
(g) MDC plot corresponding to the spectrum in the black square in (e). (h)
Comparison of the band dispersions at $\Gamma$ for the sample with different
x. (i) First-principles calculations of band structure along $\Gamma$-$M$ and
$\Gamma$-$Z$. The dashed box shows the SOC gap of the inverted bands. (j) MDC
curvature plot of the band data from ARPES, which enhances vertical bands (or
the vertical part of one band) but suppresses horizontal bands (or the
horizontal part of one band).The red dots trace the points where the intensity
of the MDC curvature exceeds the red bar in the color-scale indicator, and the
blue lines are guides to the eye indicating the band dispersion. (k) Summary
of the overall band structure.The background image is amix of raw intensity
and EDC curvature (the area in the dashed box). The bottom hole-like band is
the bulk valence band, whereas the Dirac-cone--type band is the surface band.
(l) Sketch of the spin-helical FS and the band structure along $k_{y}$, the
sample $\Gamma$$M$ direction. The EDCs at cuts 1 and 2 were measured with
SARPES. The spin pattern comes from the bottom surface. (m) Comparison of the
EDCs from SARPES and HR-ARPES measurements. The large broadening in the SARPES
measurement could be partly responsible for the small spin polarization.(n)
Spin-resolved EDCs at cut 1. (p) Spin polarization curve at cut 1. (o and q)
Same as (n) and (p), but for EDCs at cut 2. The measured spin polarizations
are consistent with the spin-helical texture illustrated in (l). }%
\label{fig2}%
\end{figure*}

The three typical materials to realize the aforementioned topological quantum
states of matter described by the three effective \textbf{k}$\cdot$\textbf{p}
Hamiltonian are monolayer FeSe/STO, monolayer FeTe$_{1-x}$Se$_{x}$/STO and
FeTeSe single crystal. To experimentally identify these topological states,
scanning tunneling microscopy/spectroscopy (STM/S) and ARPES are very powerful
tools. STM/S is a real space surface measurement technique that measures the
density of states as a function of position, and can be used to distinguish
the edge states from bulk states\cite{Yang_prl_2012,Drozdov_np_2014}. ARPES is
a momentum space measurement technique that can directly read out the band
structure, and can be used to evaluate the band evolution. The experimental
results from STM/S and ARPES for these three materials are summarized in Fig.
\ref{fig2}.

For monolayer FeSe/STO, the idea is based on comparing the gap (band gap and
superconducting gap) from dI/dV of STS with the energy distribution curve
(EDC) of ARPES in Fig. \ref{fig2} (a) to determine the bulk gap. Then, the
topological states possess the edge states, which cross the bulk band gap and
are different from the trivial normal chemical edge
states\cite{Wang_nm_2016-1}. The contribution to the density of states from
the topological edge states can be extracted by comparing the STS spectra
between the bulk regime and the edge regime. Fig. \ref{fig2} (c) and (d) are
theoretical and experimental results, respectively. The key experimental
observations are shown in Fig. \ref{fig2} (c), from which, one can find that
there exists some additional states from the edges after subtracting the
contributions from the bulk background. However, only this feature is not
enough to prove the nontrivial characteristics of the edge states. The trivial
edge states can also have similar dI/dV behaviors. In
Ref.\cite{Wang_nm_2016-1}, the checkerboard antiferromagnetic order is assumed
to exist to open a trivial gap around $M$ point in the monolayer FeSe/SrTO. In
Ref.\cite{Hao_prx_2014}, the trivial band gap at $M$ point is natural by
taking into account the tension from SrTiO$_{3}$ substrate. Furthermore, the
coexistence of antiferromagnetic order and superconducting order is doubtable
in monolayer FeSe/STO, because the gap from the antiferromagnetic order is
about 50meV, which should be easily to detect. For example, the gap should
disappear above the antiferromagnetic transition temperature T$_{N}$. Thus,
the nontrivial characteristics of the edge states should be further tested by
other experimental method such as the spin-resolved STM or nonlocal
transport\cite{Roth_science_2009}.

In Fe(Te,Se) thin films, the topological phase transition appears when
increasing the Te substitution of Se. Pictorial band evolution as change as Te
substitution is a promising evidence to testify the topological phase
transition in Fe(Te,Se) thin film. Therefore, ARPES experiment is the primary
choice. Fig. \ref{fig2} (h) summarizes the band dispersions at $\Gamma$ point
for the samples with different $x$. The experiment results show that a
down-shifting electron-like band move towards the hole-like and the band gap
between them decreases rapidly when the Se content remains shrinks. Eventually
the bands touch each other at a Se concentration of approximately 33\%, which
is further revealed in the plots of the constant energy contours and momentum
distribution curves, as shown in Fig. \ref{fig2} (e)-(g). The touch point
corresponds to the critical point of band inversion. The ARPES experimental
results give the indirect evidence for the topological band structure in
monolayer FeTe$_{1-x}$Se$_{x}$/STO\cite{Shi_sb_2017}.

For the bulk FeTe$_{1-x}$Se$_{x}$ single crystal, the emergence of electron
band No. 4 in Fig. \ref{fig2} (i) is the key ingredient to produce topological
states when increasing the Te substitution of Se. The early ARPES experiment
proved its existence through introducing the electron doping with $in$ $situ$
K evaporation\cite{Wang_prb_2015}. The newly high energy and momentum
resolution ARPES (HR-ARPES) (Energy resolution$\sim$70 $\mu$eV) and the
spin-resolved ARPES (SARPES) (Energy resolution$\sim$1.7 meV) provide powerful
tools to directly observe the topological surface states and their spin
polarization. Fig. \ref{fig2} (j) and (k) clearly demonstrate the topological
surface states with Dirac cone structure. Fig. \ref{fig2} (n)-(q) identify the
helical spin structure of the topological surface states. The combination of
HR-ARPES and SARPES results directly proved the topological band structure in
the bulk FeTe$_{1-x}$Se$_{x}$ single crystal\cite{Zhang_science_2018}.
Recently, the similar topological band structure has also been identified in
Li(Fe,Co)As\cite{Zhang_arxiv_2018-1}, which not only confirms theoretical
predictions but also proves the generic existence of tunable topological
states in iron-based superconductors.

\section{Connate topological superconductivity}

\subsection{Material proposals}

\begin{figure}[ptb]
\begin{center}
\includegraphics[width=1.0\linewidth]{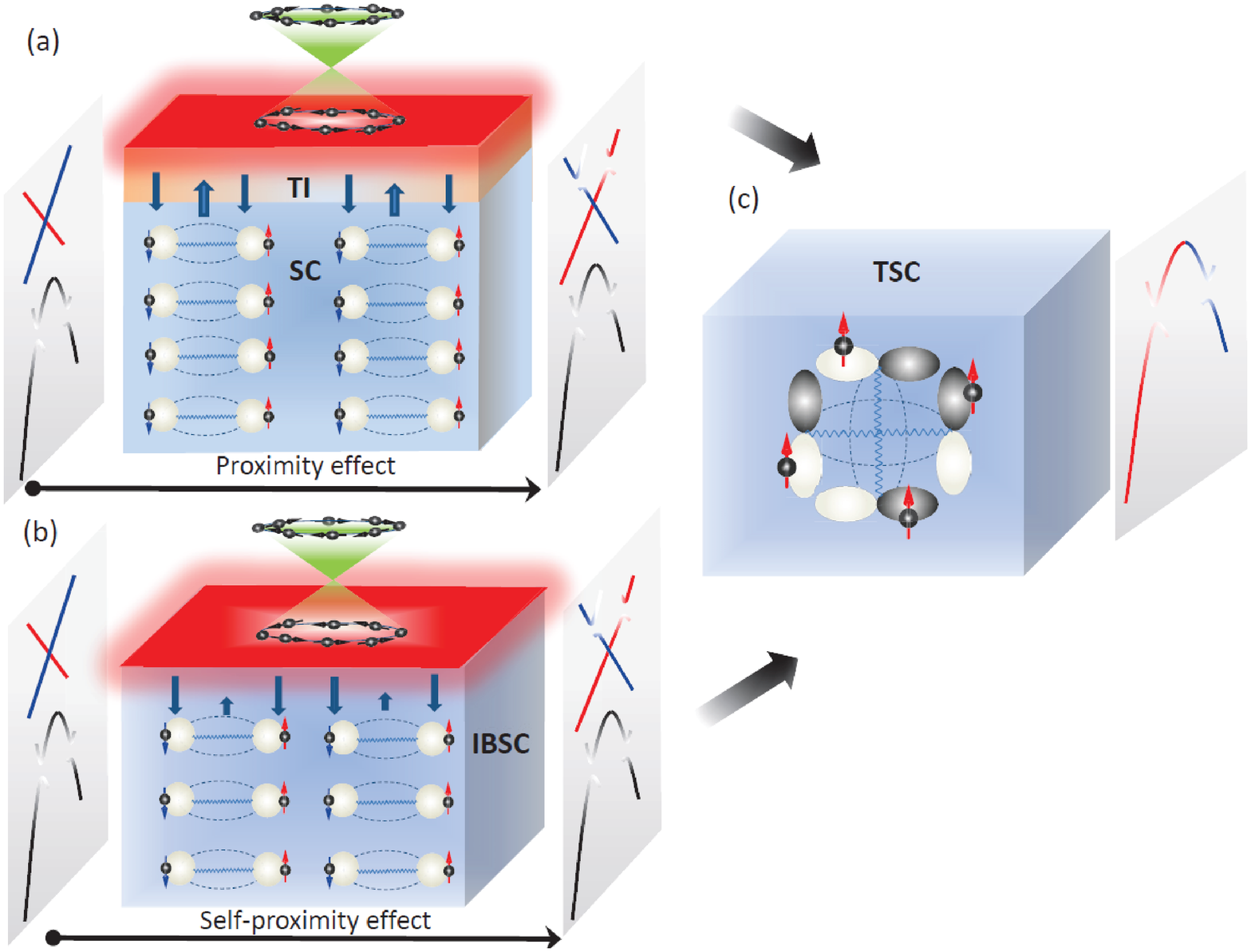}
\end{center}
\caption{(Color Online) Schematic illustrations of three kinds of strategies
to realize topological superconducting states. (a) The hetero-structure
involving conventional s-wave superconductor and topological insulator film.
(b) The iron-based superconductors with topological surface states. (c) The
unconventional superconductors with odd-parity pairing, i.e., the
spin-polarized $p+ip$ pairing here. }%
\label{fig3}%
\end{figure}

As we have mentioned in the introduction, a standard topological
superconductor requires an odd-parity pairing, as shown in Fig. \ref{fig3}
(c). The famous representative materials including Sr$_{2}$RuO$_{4}%
$\cite{Mackenzie_rmp_2003} and doped topological insulators Cu$_{x}$Bi$_{2}%
$Se$_{3}$ and Sr$_{x}$Bi$_{2}$Se$_{3}$%
\cite{Fu_prl_2010,Hor_prl_2010,Kriener_prl_2011,Sasaki_prl_2011,Levy_prl_2013,Mizushima_prb_2014,Lahoud_prb_2013,Liu_jacs_2015}
are proposed to be potential topological superconductors. However, the
experimental situation is far from definitive, because the odd-parity pairing
imposes restrictions to the pairing in spin-triplet channel, which is very
rare in solid-state materials. Therefore, the recent research mainly focuses
on some artificial structures which use the proximity effect from conventional
superconductors on the surface/edge states of the three/two-dimensional
topological insulator, on semiconductor film/nanowire with strong Rashba
spin-orbit coupling, and on iron atom
chain\cite{Fu_prl_2008,Sau_prl_2010,Lutchyn_prl_2010,Alicea_prb_2010,Mourik_science_2012,Nadj_science_2014,Albrecht_nature_2016,Xu_np_2014,Xu_prl_2015,Sun_prl_2016}%
, as shown in Fig. \ref{fig3} (a). Effectively, the model described the
structure in Fig. \ref{fig3} (a) eventually reduce into the simpler model in
Fig. \ref{fig3} (c). The ultra-low superconducting transition temperature and
the uncontrollability and uncertainty induced by the mismatch between
different materials in the artificial structures take many undetermined
problems and make these structures far beyond
practicability\cite{Xu_prl_2015,Sun_prl_2016}.

The superconductivity in iron-based superconductors is very robust against the
fine tuning the band structures. Furthermore, the aforementioned topological
phase transitions around $\Gamma$, $M$ and $\Gamma$-$Z$ line have no overall
band gap because the iron-based superconductors are multi-orbital type and
there exist other trivial bands across the Fermi energy besides the
topological bands. When the temperature decreases below the superconducting
transition temperature, the trivial bands across the Fermi energy open a
superconducting gap due the formation of the cooper pairs. At the boundaries
of the materials, the topological bands support the surface/edge states, which
also cross the Fermi energy. In comparison with trivial or extrinsic proximity
effect involving two different kinds of materials in Fig. \ref{fig3} (a), the
inducing superconductivity from trivial bulk bands to topological boundary
bands happens in a single material, and can also be called intrinsic or
self-proximity effect, as shown in Fig. \ref{fig3} (b). When the Fermi energy
is close to the surface Dirac point to guarantee the good approximation of the
linear dispersion of the surface Dirac band, the superconducting single Dirac
band can be reduced into a spinless $p_{x}+ip_{y}$
superconductor\cite{Fu_prl_2008,Xu_prl_2016-1}, which is a topological
superconductor, as shown in Fig. \ref{fig3} (c). When the $\pi$-flux vortex is
formed in the magnetic field, the effective topological superconductor can
support the zero-energy vortex-line end states, which are called Majorana modes.

Keeping the aforementioned picture in mind, one can find that all iron-based
superconductors with topological band structures can support topological
superconductors. For the monolayer FeSe/STO, the heavy hole-doped case can
support the topological edge states while the electron-doped case can support
extremely high-temperature superconductivity. Then, the boundary between the
hole-doped and electron-doped regimes in a single monolayer sample can produce
one-dimensional topological superconductor. For monolayer FeTe$_{1-x}$Se$_{x}%
$/STO, the superconductivity is robust in the whole doping
regime\cite{Shi_sb_2017}. The topological edge states emerge when $x<0.33$,
and the cooper pairs from the electron bands near $M$ point can be scattered
into the topological edge states from topological bands near $\Gamma$ point.
Then, the system spontaneously transform into the topological superconductor.
For (Ca,Pr)FeAs$_{2}$ and Ca$_{1-x}$La$_{x}$FeAs$_{2}$, the distorted As
chains in CaAs layers support topological edge states through the topological
bands near $B$ points, while the FeAs layers support superconductivity through
the trivial bulk band near both $\Gamma$ and $M$ points. The self-proximity
effect can induce the one-dimensional topological superconductivity in both
(Ca,Pr)FeAs$_{2}$ and Ca$_{1-x}$La$_{x}$FeAs$_{2}$%
\cite{Wu_prb_2014-1,Wu_prb_2015-1}. For bulk FeTe$_{1-x}$Se$_{x}$ single
crystal, the topological Dirac-cone type surface states emerge at $\bar
{\Gamma}$ point in the (001) surface Brillouin zone in the topological doped
regime. Then, the cooper pairs from the trivial bulk bands near $\Gamma$-$Z$
line and the $M$-$A$ line can be scattered into the topological Dirac-cone
type surface states. These primary and secondary self-proximity effect can
drive the bulk FeTe$_{1-x}$Se$_{x}$ single crystal into the two-dimensional
topological superconductor.

\subsection{Experiments and open questions}

For the monolayer FeSe/STO and FeTe$_{1-x}$Se$_{x}$/STO, the monolayer FeSe
and FeTe$_{1-x}$Se$_{x}$ grow on the substrate STO through the assistant of
molecular beam epitaxy (MBE). Until now, both systems have the highest
superconducting transition temperature among all iron-based superconductors,
whereas they are unstable in the air. The shortcoming takes challenges to the
devices fabrication and relevant transport measurement. On a contrary, the
bulk FeTe$_{1-x}$Se$_{x}$ single crystal is quite stable and has nice (001)
cleavage surface. More importantly, the topological superconducting states are
two-dimensional. The spontaneously generated vortex under external magnetic
field could bound Majorana zero-energy mode if the superconducting state is
topological. Then, some experimental methods like ARPES and STM/S can be used
to verify the topological superconducting state and detect the Majorana
zero-energy modes. Based on these upsides, most experimental progresses are
mainly made in the bulk FeTe$_{1-x}$Se$_{x}$ single crystal and (Li$_{0.84}%
$Fe$_{0.16}$)OHFeSe single
crystal\cite{Yin_np_2015,Zhang_science_2018,Wang_science_2018,Chen_nc_2018,Liu_arxiv_2018}%
. Along time line, we review these experiments in the following.
\begin{figure*}[ptb]
\begin{center}
\includegraphics[width=1.0\linewidth]{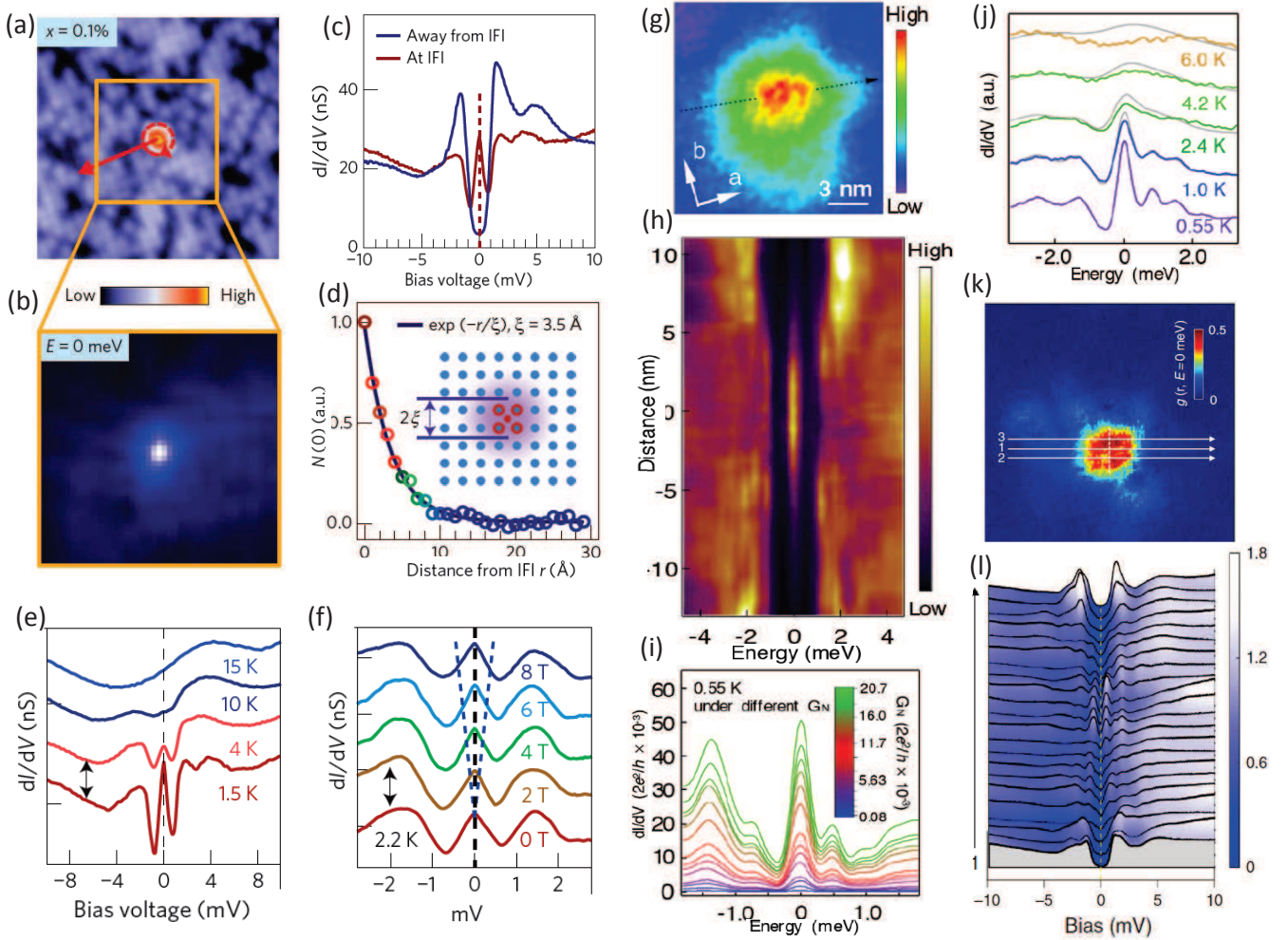}
\end{center}
\caption{(Color Online) (a)-(f) The STM/S experimental results for zero-bias
states trapped by interstitial Fe impurity in FeTe$_{0.57}$Se$_{0.43}%
$\cite{Yin_np_2015}. (g)-(l) The STM/S experimental results for bound states
trapped by vortex in FeTe$_{0.55}$Se$_{0.45}$%
\cite{Wang_science_2018,Chen_nc_2018}. (a) Topographic image of an isolated
single interstitial Fe impurity (100$\times$100\AA ). (b) Zero-energy map for
the area boxed in (a). (c) Spectra taken on top of and away from the
interstitial Fe impurity. (d) Zero-energy peak value $N(0)$ versus distance
$r$ from single interstitial Fe impurity. The solid curve is an exponential
fit with $\xi=$3.5\AA . Inset is a schematic image for the spatial
distribution of interstitial Fe impurity scattering. (e) The spectra taken at
the same interstitial Fe impurity at different temperatures. (f) The spectra
taken at the same interstitial Fe impurity under different magnetic fields.
The blue V-shaped dashed line is a guide to the eye showing the expected
Zeeman splitting ($g=2$). (g) A zero-bias conductance map (area 15nm$\times
$15nm) around vortex cores. (h) A line-cut intensity plot along the black dash
line indicated in (g). (i) Evolution of zero-bias peaks with tunneling barrier
measured at 0.55 K. $G_{N}=I_{t}/V_{s}$, which corresponds to the
energy-averaged conductance of normal states, and represents the conductance
of the tunneling barrier. $I_{t}$ and $V_{s}$ are the STS setpoint parameters.
(j) Temperature evolution of zero bias peaks in a vortex core. The gray curves
are numerically broadened 0.55 K data at each temperature. (k) Image of a
single vortex in a 20 nm$\times$20 nm region measured at 0.48 K and 4T. (l)
Tunneling spectra measured along the arrowed lines marked 1 in (k) with
increment steps of 7.6\AA . The dashed line shows the position of zero bias
voltage. The discrete CdGM bound state peaks can be clearly observed near the
vortex core center.}%
\label{fig4}%
\end{figure*}

The first unexpected experiment is about the impurity bound states in
FeTe$_{0.57}$Se$_{0.43}$ single crystals\cite{Yin_np_2015}. FeTe$_{0.57}%
$Se$_{0.43}$ single crystals contain a large amount of excess iron that as
single iron atoms randomly situate at the interstitial sites between two (Te,
Se) atomic planes\cite{Taen_prb_2009}. The STM/S spectrum observed a strong
zero-energy bound state at the centre of the single interstitial Fe impurity.
The experimental results are summarized in Fig. \ref{fig4} (a)-(f). The
zero-energy bound state has the following features. (1) The spatial pattern of
the zero-energy bound state is almost circular, which is different from the
cross-shape pattern of the Zn impurity in Bi$_{2}$Sr$_{2}$Ca(Cu,Zn)$_{2}%
$O$_{8+\delta}$\cite{Pan_nature_2000}. (2) The intensity of the zero-energy
bound state exponentially decays with a characteristic length of $\xi
=3.5$\AA , which is almost one order of magnitude smaller than the typical
coherent length of $25$\AA \ in the iron-based
superconductor\cite{Yin_prl_2009,Shan_np_2011}. (3) The bound state is
strictly at zero even the external magnetic field increases to 8T. (4) The
zero-energy bound state peak remains at zero energy even when two interstitial
Fe impurity atoms are located near each other ($\sim15$\AA ). It is a serious
challenge to consistently explain these features of the zero-energy bound
state induced by interstitial Fe impurity. The d-wave pairing symmetry
scenario can result in a zero-energy bound state at unitary
limit\cite{Balatsky_rmp_2006}, but violates feature (1). The Kondo impurity
resonance scenario can give an accidental zero-energy bound
state\cite{Balatsky_rmp_2006}, but violates feature (4). A fascinating
scenario is that the mode is Majorana zero-energy
mode\cite{Read_prb_2000,Kitaev_pu_2001}, which captures features (1)-(3).
Recently, a theoretical work claimed that an interstitial Fe impurity could
bound an quantum anomalous vortex without magnetic field, and the quantum
anomalous vortex can bound a Majorana zero-energy mode when topological
surface states of FeTe$_{0.57}$Se$_{0.43}$ become
superconducting\cite{Jiang_arxiv_2018}. However, it is still hard to explain
feature (4) by the Majorana zero-energy mode scenario. Until now, the origin
of the zero-energy bound state trapped by interstitial Fe impurity is still
underdetermined. Topological or other reasons need further experimental and
theoretical explorations.

The second experimental breakthrough is about the vortex bound states on the
surface of FeTe$_{0.55}$Se$_{0.45}$ single
crystals\cite{Wang_science_2018,Chen_nc_2018}. FeTe$_{0.55}$Se$_{0.45}$
belongs to type II superconductor. Once a small external magnetic field is
applies along c-axis, magnetic vortex structures are formed due to the small
lower critical field $H_{c1}$. The high-resolution STM/S can measure the bound
states trapped by the vortex. Two experimental group claimed completely
different results for the same material FeTe$_{0.55}$Se$_{0.45}$ single
crystals. The former group claimed that they observed a sharp zero-bias peak
inside a vortex core that does not split when moving away from the vortex
center, which could be attributed to the nearly pure Majorana bound
state\cite{Wang_science_2018}. The experimental results are summarized in Fig.
\ref{fig4} (g)-(j). The vortex bound states exhibit the following features.
(1) Statistically, there are about 20\% success rate in observing the isolated
pure Majorana bound state during more than 150 measurements. (2) Across a
large range of magnetic fields the observed zero-bias peak does not split when
moving away from a vortex center. (3) Most of the observed zero-bias peak
vanish around 3K. (4) Robust zero-bias peaks can be observed over two orders
of magnitude in tunneling barrier conductance, with the width barely changing.
Feature (1) is argued to attribute to the disorder effect and/or inhomogeneous
distribution of Te/Se. Feature (2) is attributed to the large $\Delta
_{sc}/E_{F}$ ratio in this system. Feature (3) is attributed to that the
Caroli-de-Gennes-Matricon (CdGM) state\cite{Caroli_pl_1964} is protected by a
mini-energy gap with with a temperature about $\Delta_{sc}^{2}/E_{F}\sim3$K,
and the thermal excitation around and beyond 3K can kill the CdGM state.
Feature (4) indicated the line width of zero-bias peaks is almost completely
limited by the combined broadening of energy resolution and STM thermal
effect, suggesting that the intrinsic width of the Majorana bound state is
much smaller in the weak tunnelling
regime\cite{Setiawan_prb_2017,Colbert_prb_2014}. The detailed experimental
measurements eliminate some scenarios to cause a zero-bias peak in tunneling
experiments, such as antilocalization, reflectionless tunneling, Kondo effect,
Josephson supercurrent and packed CdGM states near zero
energy\cite{Pikulin_njp_2012,Bagrets_prl_2012,Wees_prl_1992,Eduardo_prl_2012,Churchill_prb_2013,Levy_prl_2013,Hess_prl_1990,Gygi_prb_1990}%
. Features (2)-(4) can be well understood with the Majorana bound state
scenario, it is probable that the observed zero-bias peaks correspond to
Majorana bound state. However, the feature (1) is a serious problem, which is
different from other proposals to realize Majorana bound states. In the
present experiments, it seems no comprehensive evidences of the disorder
effect and/or influence of inhomogeneous distribution of Te/Se are provided.
Furthermore, if the observed zero-bias peaks are from Majorana bound states,
the non-Abelian statistics can be demonstrated by move a vortex with a STM
tip. This kind of experiment is the smoking gun for Majorana modes. Another
experimental group claimed that they only observed the trivial CdGM bound
state trapped by vortex in the same FeTe$_{0.55}$Se$_{0.45}$ single crystals.
For statistics, the energies of bound state peaks close to the zero bias\ are
collected from all measured nine vortices presented\cite{Chen_nc_2018}. The
experimental results are summarized in Fig. \ref{fig4} (k)-(l). In principle,
there should be a special vortex to bound the zero-bias peak according to the
20\% success rate claimed in the former experiment. Unfortunately, two
experiments for the same material from two groups give the inconsistent
results\cite{Wang_science_2018,Chen_nc_2018}. The argument about the
difference being attributed to the different annealed process is not very
convincing. It seems that the appearance of zero-bias peaks is selective. The
behaviors challenge the topological origin, which is usually universal and robust.

The third subsequent experiment is about the vortex bound states on the FeSe
cleavage plane of (Li$_{0.84}$Fe$_{0.16}$)OHFeSe single
crystal\cite{Liu_arxiv_2018}. In compared with FeTe$_{0.55}$Se$_{0.45}$, the
superconducting FeSe layers in (Li$_{0.84}$Fe$_{0.16}$)OHFeSe are
stoichiometric. Therefore, there exist defect-free areas, which support the
un-pinned or free vortex. The STM/S measurements show that (1) The free vortex
cores bound zero-bias modes, which do not shift with varying underlying
superconducting gap as the other peaks do. (2) the zero-bias modes survive to
high magnetic field due to the short coherence length. (3) The zero-bias mode
coexists with other low-lying CdGM states but separates from each other. These
features are similar to those of the zero-bias modes observed in FeTe$_{0.55}%
$Se$_{0.45}$. Therefore, the zero-bias modes can be also attributed to
Majorana zero-energy modes, and can be argued to have topological origin in
(Li$_{0.84}$Fe$_{0.16}$)OHFeSe. However, the topological origin in
(Li$_{0.84}$Fe$_{0.16}$)OHFeSe is underdetermined, unlike FeTe$_{0.55}%
$Se$_{0.45}$ with solid experimental evidences for the topological band
structure. Recall the discussions about the band inversion along $\Gamma$-$Z$
line in FeTe$_{0.55}$Se$_{0.45}$ in Section II A, the strong dispersion of
band 4 in Fig. \ref{fig2} (j) benefits from the quite small layer distance and
large size of Te atoms. The band 4 in pure FeSe is flat\cite{Wang_prb_2015}.
It is very strange that the band 4 in (Li$_{0.84}$Fe$_{0.16}$)OHFeSe has
strong dispersion. Furthermore, the band gap open is due to the strong
spin-orbit coupling from Te atom not Se atom. Another critical condition to
obtain the topological surface states is that the chemical potential must
properly lie in the quite small band gap. However, the chemical potential in
(Li$_{0.84}$Fe$_{0.16}$)OHFeSe is far from the band gap. In this situation,
the top and bottom surfaces start to communicate with each other and break the
zero-bias mode. At last, it lacks the smoking gun ARPES experiment to prove
the helical structure of the claimed observed topological surface states in
(Li$_{0.84}$Fe$_{0.16}$)OHFeSe. In summary, the experimental observations of
the zero-bias modes in (Li$_{0.84}$Fe$_{0.16}$)OHFeSe is more clear, but the
topological origin needs to be understood.

\section{Summary and Perspectives}

The discovery of topological insulators has established a standard paradigm to
guide the communities to pursue the topological states of matter in quantum
materials. Such pursuits cause intersections between the topology and
iron-based superconductors. As emphasized in this review, important principles
for the theoretical understandings of the energy-band topology in new
materials include applying general concepts with the help of symmetry analysis
and constructing the effective models. For the iron-based superconductors, the
multi-orbital band structures and the diversity of materials provide
opportunities to realize the effective theoretical models. These topological
materials include monolayer FeSe/STO, monolayer Fee$_{1-x}$Se$_{x}$/STO,
FeTe$_{1-x}$Se$_{x}$, and LiFe$_{1-x}$Co$_{x}$As etc.

In the superconducting states, it is naturally expected to obtain the
topological superconducting states with the help of self-proximity effect.
However, different from the energy-band topology, the expected topological
superconducting states exhibit many unexpected experimental phenomena,
including the surprising robust zero-energy mode trapped by Fe impurity in
FeTe$_{0.57}$Se$_{0.43}$, the selective appearance of zero-bias mode trapped
by vortex in FeTe$_{0.57}$Se$_{0.43}$, and the coexistence of zero-bias mode
and CdGM states trapped by free vortex in (Li$_{0.84}$Fe$_{0.16}$)OHFeSe. Even
if all these phenomena are attributed to Majorana zero-energy modes, there are
deep inconsistencies within different experiments as well as between
experiments and theories. In this respect, clarifying the creation mechanism
of these so-called Majorana zero-energy modes are worth pursuing. For such
efforts, the availability of the high-quality single crystal, whose chemical
potential can be artificially fine tuned, would be crucial. Once the physics
of the so-called Majorana zero-energy modes are clarified, finding ways to
manipulate the non-Abelian statistics of the Majorana zero-energy modes is a
significant challenge for future applications in quantum computing.

Iron-based superconductor owns a rich phase diagram. Beside the normal and
superconducting phases, there include nematic phase, orbital ordering phase
and various antiferromagnetic phases. Searching the topology embedded in these
ordered phases would be interesting. For the theoretical aspect, there have
been some studies\cite{Wu_arxiv_2016,Hao_prb_2017}, but the experimental
exploration is blank. In future, a stronger collaboration between theory and
experiment is required to explore topological quantum states in new materials
of iron-based superconductors.

Finally, it can not be entirely ruled out that the superconducting states of
iron-based superconductors themselves could be highly unconventional. In the
ten years of the research of iron-based superconductors, there still are many
unsolved puzzles\cite{Hirschfeld_rpp_2011,Chubukov_arxiv_2014} observed by a
variety of different experimental methods, such as transport, Raman spectrum,
neutron scattering, nuclear magnetic resonance, electron spin resonance,
STM/S, and ARPES etc. For example,  the  interplay between spin, orbital, lattice and charge degrees of freedom  is not fully understood,  not only is there no smoking gun proof for the s$_{\pm}$ pairing yet but also it is clear that the s$_{\pm}$ pairing symmetry cannot be valid for many iron-chalcogenide systems, whether there is a sign change in the superconducting states of iron-chalcogenide systems without hole pockets  or not are highly debated, and  the origin of the enhancement of  the transition
temperature  found in single layer FeSe remains to be understood.  The topological exploration in iron-based
superconductors may help us to discover surprising characters and mechanism
hidden behind superconducting pairing, and leads to answers to these
unsolved puzzles.

\begin{acknowledgments}
N. Hao thanks the KITS, UCAS for the hospitality during his visit. This work
was supported by the National Key R\&D Program of China (No.2015CB921300,
No.2017YFA0303100, 2017YFA0303201),National Natural Science Foundation of
China (No. 11674331, No. 11334012), the `100 Talents Project' of Chinese
Academy of Sciences, and the Strategic Priority Research Program of CAS (No.XDB07000000).
\end{acknowledgments}

%\bibliographystyle{plain}
%\bibliographystyle{abbrv}
%\bibliography{NSR_ref}

\begin{thebibliography}{99}                                                                                               %


\bibitem {Hasan_rmp_2010}Hasan MZ and Kane CL. \newblock Colloquium:
topological insulators. \newblock {\em Rev Mod Phys} 2010; \textbf{82}:3045.

\bibitem {Qi_rmp_2011}Qi XL and Zhang SC. \newblock Topological insulators and
superconductors. \newblock {\em Rev Mod Phys} 2011; \textbf{83}:1057.

\bibitem {Armitage_rmp_2018}Armitage NP, Mele EJ and Vishwanath A. \newblock
Weyl and Dirac semimetals in three-dimensional solids.
\newblock {\em Rev Mod Phys} 2018; \textbf{90}:015001.

\bibitem {Nayak_rmp_2008}Nayak C, Simon SH and Stern A \textit{et al}.
\newblock Non-abelian anyons and topological quantum computation.
\newblock {\em Rev Mod Phys} 2008; \textbf{80}:1083.

\bibitem {Fu_prl_2008}Fu L and Kane CL. \newblock Superconducting proximity
effect and majorana fermions at the surface of a topological insulator.
\newblock {\em Phys Rev Lett} 2008; \textbf{100}:096407.

\bibitem {Lutchyn_prl_2010}Lutchyn RM, Sau JD and Sarma SD. \newblock Majorana
fermions and a topological phase transition in semiconductor-superconductor
heterostructures. \newblock {\em Phys Rev Lett} 2010; \textbf{105}:077001.

\bibitem {Sau_prl_2010}Sau JD, Lutchyn, RM and Tewari S \textit{et al}.
\newblock Generic new platform for topological quantum computation using
semiconductor heterostructures. \newblock {\em Phys Rev Lett} 2010;
\textbf{104}:040502.

\bibitem {Stevan_science_2014}Nadj-Perge S, Drozdov IK and Li J \textit{et
al}. \newblock Observation of majorana fermions in ferromagnetic atomic chains
on a superconductor. \newblock {\em Science} 2014, 10.1126/science.1259327.

\bibitem {Shi_sb_2017}Shi X, Han ZQ, Richard P \textit{et al}. \newblock
FeTe$_{1-x}$Se$_{x}$ monolayer films: towards the realization of
high-temperature connate topological superconductivity.
\newblock {\em Science bulletin} 2017; \textbf{62}:503--507.

\bibitem {Xu_prl_2016}Xu G, Lian B and Tang P \textit{et al}. \newblock
Topological superconductivity on the surface of Fe-based superconductors.
\newblock {\em Physl Rev Lett} 2016; \textbf{117}:047001.

\bibitem {Hao_prx_2014}Hao N and Hu J. \newblock Topological phases in the
single-layer FeSe. \newblock {\em Phys Rev X} 2014; \textbf{4}:031053.

\bibitem {Wu_prb_2016}Wu X, Qin S and Liang Y \textit{et al}. \newblock
Topological characters in $\mathrm{Fe}(\mathrm{T}{\mathrm{e}}_{1-x}%
\mathrm{S}{\mathrm{e}}_{x})$ thin films. \newblock {\em Phys Rev B} 2016;
\textbf{93}:115129.

\bibitem {Wang_prb_2015}Wang Z, Zhang P and Xu G \textit{et al.} \newblock
Topological nature of the $\mathrm{F}{\mathrm{eSe}}_{0.5}\mathrm{T}%
{\mathrm{e}}_{0.5}$ superconductor. \newblock {\em Phys Rev B} 2015;
\textbf{92}:115119.

\bibitem {Zhang_arxiv_2018-1}Zhang P, Wu X and Yaji K \textit{et al}.
\newblock Direct observation of multiple topological phases in the iron-based
superconductor Li(Fe,Co)As.
\newblock {\em Nature Physics (online), arXiv:1803.00846} 2018.

\bibitem {Wu_prb_2015-1}Wu X, Qin S and Liang Y \textit{et al}.
\newblock$\mathrm{C}{\mathrm{aFeAs}}_{2}$: A staggered intercalation of
quantum spin Hall and high-temperature superconductivity.
\newblock {\em Phys Rev B} 2015; \textbf{91}:081111.

\bibitem {Zhang_science_2018}Zhang P, Yaji K and Hashimoto T \textit{et al}.
\newblock Observation of topological superconductivity on the surface of an
iron-based superconductor. \newblock {\em Science} 2018; \textbf{360}:182--186.

\bibitem {Wang_science_2018}Wang D, Kong L and Fan P \textit{et~al}.
\newblock
Evidence for majorana bound states in an iron-based superconductor.
\newblock {\em Science} 2018; \textbf{362}:333-335.

\bibitem {Liu_arxiv_2018}Liu Q, Chen C and Zhang T \textit{et al.} \newblock
Robust and clean majorana zero mode in the vortex core of high-temperature
superconductor (Li$_{0.84}$Fe$_{0.16}$)OHFeSe.
\newblock {\em arXiv:1807.01278 [cond-mat.supr-con]}, 2018.

\bibitem {Paglione_np_2010}Paglione J and Greene RL. \newblock
High-temperature superconductivity in iron-based materials.
\newblock {\em Nat phys} 2010; \textbf{6}:645.

\bibitem {Guterding_prb_2017}Guterding D, Jeschke HO, and Valent\'{\i} R.
\newblock Basic electronic properties of iron selenide under variation of
structural parameters. \newblock {\em Phys Rev B} 2017; \textbf{96}:125107.

\bibitem {Singh_prl_2008}Singh DJ and Du MH. \newblock Density functional
study of LaFeAsO$_{1-x}$F$_{x}$: A low carrier density superconductor near
itinerant magnetism. \newblock {\em Phys Rev Lett} 2008; \textbf{100}:237003.

\bibitem {Singh_prb_2008}Singh DJ. \newblock Electronic structure and doping
in BaFe$_{2}$As$_{2}$ and lifeas: Density functional calculations.
\newblock {\em Phys Rev B} 2008; \textbf{78}:094511.

\bibitem {Podolsky-1}Podolsky D. Untangling the orbitals in iron-based
superconductors. \textit{Physics }2012; \textbf{5}:61.

\bibitem {Liu_prb_2010}Liu C, Qi X and Zhang H \textit{et al}. \newblock Model
Hamiltonian for topological insulators. \newblock {\em Phys Rev B} 2010;
\textbf{82}:045122.

\bibitem {Hao-2018}Hao N and J. Hu. Research progress of topological quantum
states in iron-based superconductor. \textit{Acta Phys Sin} 2018; \textbf{67}:207101.

\bibitem {Kane_prl_2005-1}Kane CL and Mele EJ. \newblock${Z}_{2}$ topological
order and the quantum spin Hall effect. \newblock {\em Phys Rev Lett} 2005;
\textbf{95}:146802.

\bibitem {Bernevig_science_2006}Bernevig BA, Hughes TL and Zhang S.
\newblock
Quantum spin Hall effect and topological phase transition in HgTe quantum
wells. \newblock {\em Science} 2006; \textbf{314}:1757--1761.

\bibitem {Konig_science_2007}K\H{o}nig M, Wiedmann S and Br\H{u}ne C
\textit{et al}. \newblock Quantum spin Hall insulator state in HgTe quantum
wells. \newblock {\em Science} 2007; \textbf{318}:766--770.

\bibitem {Moore_prb_2007}Moore JE and Balents L. \newblock Topological
invariants of time-reversal-invariant band structures.
\newblock {\em Phys Rev B} 2007; \textbf{75}:121306.

\bibitem {Fu_prl_2007}Fu L, Kane CL, and Mele EJ. \newblock Topological
insulators in three dimensions. \newblock {\em Phys Rev Lett} 2007;
\textbf{98}:106803.

\bibitem {Roy_prb_2009}Roy R. \newblock Topological phases and the quantum
spin Hall effect in three dimensions. \newblock {\em Phys Rev B} 2009;
\textbf{79}:195322.

\bibitem {Fu_prb_2007}Fu L and Kane CL. \newblock Topological insulators with
inversion symmetry. \newblock {\em Phys Rev B} 2007; \textbf{76}:045302.

\bibitem {Chen_science_2009}Chen Y, Analytis JG, Chu J \textit{et al}.
\newblock Experimental realization of a three-dimensional topological
insulator, Bi$_{2}$Te$_{3}$. \newblock {\em Science} 2009; \textbf{325}:178--181.

\bibitem {Zhang_np_2009}Zhang H, Liu C and Qi X \textit{et al}. \newblock
Topological insulators in Bi$_{2}$Se$_{3}$, Bi$_{2}$Te$_{3}$ and Sb$_{2}%
$Te$_{3}$ with a single Dirac cone on the surface. \newblock {\em Nat phys}
2009; \textbf{5}:438.

\bibitem {Xia_np_2009}Xia Y, Qian D and Hsieh D \textit{et al}. \newblock
Observation of a large-gap topological-insulator class with a single Dirac
cone on the surface. \newblock {\em Nat phys} 2009; \textbf{5}:398.

\bibitem {Cvetkovic_prb_2013}Cvetkovic V and Vafek O. \newblock Space group
symmetry, spin-orbit coupling, and the low-energy effective Hamiltonian for
iron-based superconductors. \newblock {\em Phys Rev B} 2013 \textbf{88}:134510.

\bibitem {Hao_prb_2014}Hao N and Hu J. \newblock Odd parity pairing and
nodeless antiphase ${S}_{\pm}$ in iron-based superconductors.
\newblock {\em Phys Rev B} 2014; \textbf{89}:045144.

\bibitem {Hao_prb_2015}Hao N and Shen S. \newblock Topological superconducting
states in monolayer FeSe/SrTiO$_{3}$. \newblock {\em Phys Rev B} 2015;
\textbf{92}:165104.

\bibitem {Wang_cpl_2012}Wang Q, Li Z and Zhang W \textit{et al}. \newblock
Interface-induced high-temperature superconductivity in single unit-cell FeSe
films on SrTiO$_{3}$. \newblock {\em Chin Phys Lett} 2012; \textbf{29}:037402, .

\bibitem {He_nm_2013}He S, He J and Zhang W \textit{et al}. \newblock Phase
diagram and electronic indication of high-temperature superconductivity at 65
k in single-layer FeSe films. \newblock {\em Nat Mat} 2013; \textbf{12}:605.

\bibitem {Tan_nm_2013}Tan S, Zhang Y and Xia M \textit{et al}. \newblock
Interface-induced superconductivity and strain-dependent spin density waves in
FeSe/SrTiO$_{3}$ thin films. \newblock {\em Nat Mat} 2013; \textbf{12}:634.

\bibitem {Liu_nc_2012}Liu D, Zhang W and Mou D \textit{et al}.\newblock
Electronic origin of high-temperature superconductivity in single-layer FeSe
superconductor. \newblock {\em Nat Commun} 2012; \textbf{3}:931.

\bibitem {Lee_nature_2014}Lee JJ, Schmitt FT and Moore RG \textit{et al}.
\newblock Interfacial mode coupling as the origin of the enhancement of Tc in
FeSe films on SrTiO$_{3}$. \newblock {\em Nature} 2014; \textbf{515}:245.

\bibitem {Miyata_nm_2015}Miyata Y, Nakayama K and Sugawara K \textit{et al}.
\newblock High-temperature superconductivity in potassium-coated multilayer
FeSe thin films. \newblock {\em Nat Mat} 2015 \textbf{14}:775.

\bibitem {Peng_nc_2014}Peng R, Xu HC and Tan SY \textit{et al}. \newblock
Tuning the band structure and superconductivity in single-layer FeSe by
interface engineering. \newblock {\em Nat Commun} 2014; \textbf{5}:5044.

\bibitem {Zhang_arxiv_2018}Zhang P, Wang Z and Ishida Y \textit{et al}.
\newblock Topological Dirac semimetal phase in the iron-based superconductor
Fe(Te,Se). \newblock {\em arXiv:1803.00845 [cond-mat.supr-con]}, 2018.

\bibitem {Wang_nm_2016-1}ZF~Wang, Zhang H and Liu D \textit{et al}.
\newblock
Topological edge states in a high-temperature superconductor FeSe/SrTiO$_{3}$
(001) film. \newblock {\em Nat. Mater.} 2016; \textbf{15}:968.

\bibitem {Yang_prl_2012}Yang F, Miao L and Wang ZF \textit{et al}. \newblock
Spatial and energy distribution of topological edge states in single Bi(111)
bilayer. \newblock {\em Physl Rev Lett}, 109(1):016801, 2012.

\bibitem {Drozdov_np_2014}Drozdov IK, Alexandradinata A and Jeon S \textit{et
al}. \newblock One-dimensional topological edge states of bismuth bilayers.
\newblock {\em Nat Phys}, 10:664, 2014.

\bibitem {Roth_science_2009}Roth A, Br{\"{u}}ne C and Buhmann H \textit{et
al}. \newblock Nonlocal transport in the quantum spin Hall state.
\newblock {\em Science} 2009; \textbf{325}:294--297.

\bibitem {Mackenzie_rmp_2003}Mackenzie AP and Maeno Y. \newblock The
superconductivity of Sr$_{2}$RuO$_{4}$ and the physics of spin-triplet
pairing. \newblock {\em Rev Mod Phys}, 2003; \textbf{75}:657--712.

\bibitem {Fu_prl_2010}Fu L and Berg E. \newblock Odd-parity topological
superconductors: theory and application to Cu$_{x}$Bi$_{2}$Se$_{3}$.
\newblock {\em Phys Rev Lett} 2010; \textbf{105}:097001.

\bibitem {Hor_prl_2010}Hor YS, Williams AJ and Checkelsky JG \textit{et al}.
\newblock Superconductivity in Cu$_{x}$Bi$_{2}$Se$_{3}$ and its implications
for pairing in the undoped topological insulator.
\newblock {\em Phys Rev Lett} 2010; \textbf{104}:057001.

\bibitem {Kriener_prl_2011}Kriener M, Segawa K and Ren Z \textit{et al}.
\newblock Bulk superconducting phase with a full energy gap in the doped
topological insulator Cu$_{x}$Bi$_{2}$Se$_{3}$. \newblock {\em Phys Rev Lett}
2011; \textbf{106}:127004.

\bibitem {Sasaki_prl_2011}Sasaki S, Kriener M and Segawa K \textit{et al}.
\newblock Topological superconductivity in Cu$_{x}$Bi$_{2}$Se$_{3}$.
\newblock {\em Phys Rev Lett} 2011; \textbf{107}:217001.

\bibitem {Levy_prl_2013}Levy N, Zhang T and Ha J \textit{et al}. \newblock
Experimental evidence for $s$-wave pairing symmetry in superconducting
Cu$_{x}$Bi$_{2}$Se$_{3}$ single crystals using a scanning tunneling
microscope. \newblock {\em Phys Rev Lett} 2013; \textbf{110}:117001.

\bibitem {Mizushima_prb_2014}Mizushima T , Yamakage A and Sato M \textit{et
al}. \newblock Dirac-fermion-induced parity mixing in superconducting
topological insulators. \newblock {\em Phys Rev B} 2014; \textbf{90}:184516.

\bibitem {Lahoud_prb_2013}Lahoud E, Maniv E and Petrushevsky MS et al.
\newblock Evolution of the fermi surface of a doped topological insulator with
carrier concentration. \newblock {\em Phys Rev B} 2013; \textbf{88}:195107.

\bibitem {Liu_jacs_2015}Liu Z, Yao X and Shao J \textit{et al}. \newblock
Superconductivity with topological surface state in Sr$_{x}$Bi$_{2}$Se$_{3}$.
\newblock {\em J Am Chem Soc} 2015; \textbf{137}:10512--10515.

\bibitem {Alicea_prb_2010}Alicea J. \newblock Majorana fermions in a tunable
semiconductor device. \newblock {\em Phys Rev B} 2010; \textbf{81}:125318.

\bibitem {Mourik_science_2012}Mourik V, Zuo K and Frolov SM \textit{et al}.
\newblock Signatures of majorana fermions in hybrid
superconductor-semiconductor nanowire devices. \newblock {\em Science} 2012;
\textbf{336}:1003--1007.

\bibitem {Nadj_science_2014}Nadj-Perge S, Drozdov IK and Li J et al.,
\newblock Observation of majorana fermions in ferromagnetic atomic chains on a
superconductor. \newblock {\em Science} 2014 \textbf{346}:602--607.

\bibitem {Albrecht_nature_2016}Albrecht SM, Higginbotham AP and Madsen M
\textit{et al}. \newblock Exponential protection of zero modes in Majorana
islands. \newblock {\em Nature} 2016; \textbf{531}:206.

\bibitem {Xu_np_2014}Xu S, Alidoust N and Belopolski I \textit{et al},
\newblock Momentum-space imaging of cooper pairing in a half-Dirac-gas
topological superconductor. \newblock {\em Nat Phys} 2014; \textbf{10}:943.

\bibitem {Xu_prl_2015}Xu J, Wang M and Liu Z et al. \newblock Experimental
detection of a majorana mode in the core of a magnetic vortex inside a
topological insulator-superconductor Bi$_{2}$Te$_{3}$/NbSe$_{2}$
heterostructure. \newblock {\em Phys Rev Lett} 2015; \textbf{114}:017001.

\bibitem {Sun_prl_2016}Sun H, Zhang K and Hu L \textit{et al}. \newblock
Majorana zero mode detected with spin selective Andreev reflection in the
vortex of a topological superconductor. \newblock {\em Phys Rev Lett} 2016;
\textbf{116}:257003.

\bibitem {Xu_prl_2016-1}Xu G, Lian B and Tang P \textit{et al}. \newblock
Topological superconductivity on the surface of Fe-based superconductors.
\newblock {\em Phys Rev Lett} 2016; \textbf{117}:047001.

\bibitem {Wu_prb_2014-1}Wu X, Le C and Liang Y \textit{et al}. \newblock
Effect of As-chain layers in C${\text{aFeAs}}_{2}$. \newblock {\em Phys Rev B}
2014; \textbf{89}:205102.

\bibitem {Yin_np_2015}~Yin JX, Wu Z and Wang JH \textit{et al}. \newblock
Observation of a robust zero-energy bound state in iron-based superconductor
Fe(Te,Se). \newblock {\em Nat Phys}, 11(7):543, 2015.

\bibitem {Chen_nc_2018}Chen M, Chen X and Yang H et al. \newblock Discrete
energy levels of Caroli-de Gennes-Matricon states in quantum limit in
FeTe$_{0.55}$Se$_{0.45}$. \newblock {\em Nat Commun} 2018; \textbf{9}:970.

\bibitem {Taen_prb_2009}Taen T, Tsuchiya Y and Nakajima Y \textit{et al}.
\newblock Superconductivity at $T_{c}\symbol{126}$14K in single-crystalline
FeTe$_{0.61}$Se$_{0.39}$. \newblock {\em Phys Rev B} 2009; \textbf{80}:092502.

\bibitem {Pan_nature_2000}Pan SH, Hudson EW and Lang KM \textit{et al}.
\newblock Imaging the effects of individual Zinc impurity atoms on
superconductivity in Bi$_{2}$Sr$_{2}$CaCu$_{2}$O$_{8+\delta}$.
\newblock {\em Nature} 2000; \textbf{403}:746.

\bibitem {Yin_prl_2009}Yin Y, Zech M and Williams TL \newblock Scanning
tunneling spectroscopy and vortex imaging in the iron pnictide superconductor
BaFe$_{1.8}$Co$_{0.2}$As$_{2}$. \newblock {\em Phys Rev Lett} 2009;
\textbf{102}:097002.

\bibitem {Shan_np_2011}Shan L, Wang Y and Shen B \textit{et al}. \newblock
Observation of ordered vortices with andreev bound states in Ba$_{0.6}%
$K$_{0.4}$Fe$_{2}$As$_{2}$. \newblock {\em Nat Phys} 2011; \textbf{7}:325.

\bibitem {Balatsky_rmp_2006}Balatsky AV, Vekhter I and Zhu J. \newblock
Impurity-induced states in conventional and unconventional superconductors.
\newblock {\em Rev Mod Phys} 2006; \textbf{78}:373.

\bibitem {Read_prb_2000}Read N and Green D. \newblock Paired states of
fermions in two dimensions with breaking of parity and time-reversal
symmetries and the fractional quantum Hall effect. \newblock {\em Phys Rev B}
2000; \textbf{61}:10267--10297.

\bibitem {Kitaev_pu_2001}Kitaev AY. \newblock Unpaired Majorana fermions in
quantum wires. \newblock {\em Phys Usp} 2001; \textbf{44}:131.

\bibitem {Jiang_arxiv_2018}Jiang K, Dai X and Wang Z. \newblock Quantum
anomalous vortex and majorana zero mode in Fe (Te, Se) superconductors.
\newblock {\em arXiv preprint arXiv:1808.07072}, 2018.

\bibitem {Caroli_pl_1964}Caroli C, Gennes PGD and Matricon J. \newblock Bound
fermion states on a vortex line in a type II superconductor.
\newblock {\em Phys Lett}, 9:307--309, 1964.

\bibitem {Setiawan_prb_2017}Setiawan F, Liu C and Sau JD \textit{et al}.
\newblock Electron temperature and tunnel coupling dependence of zero-bias and
almost-zero-bias conductance peaks in Majorana nanowires.
\newblock {\em Phys Rev B} 2017; \textbf{96}:184520.

\bibitem {Colbert_prb_2014}Colbert JR and Lee PA. \newblock Proposal to
measure the quasiparticle poisoning time of Majorana bound states.
\newblock {\em Phys Rev B} 2014; \textbf{89}:140505.

\bibitem {Pikulin_njp_2012}Pikulin DI, Dahlhaus JP and Wimmer M \textit{et
al}. \newblock A zero-voltage conductance peak from weak antilocalization in a
Majorana nanowire. \newblock {\em New J Phys} 2012; \textbf{14}:125011.

\bibitem {Bagrets_prl_2012}Bagrets D and Altland A. \newblock Class \textit{D}
spectral peak in Majorana quantum wires. \newblock {\em Phys Rev Lett} 2012;
\textbf{109}:227005.

\bibitem {Wees_prl_1992}Wees BJV, Vries PD and Magn\'{e}e P \textit{et al}.
\newblock Excess conductance of superconductor-semiconductor interfaces due to
phase conjugation between electrons and holes. \newblock {\em Phys Rev Lett}
1992; \textbf{69}:510--513.

\bibitem {Eduardo_prl_2012}Lee EJH, Jiang X and Aguado R \textit{et al}.
\newblock Zero-bias anomaly in a nanowire quantum dot coupled to
superconductors. \newblock {\em Phys Rev Lett} 2012; \textbf{109}:186802.

\bibitem {Churchill_prb_2013}Churchill HOH, Fatemi V and Grove-Rasmussen K
\textit{et al}. \newblock Superconductor-nanowire devices from tunneling to
the multichannel regime: Zero-bias oscillations and magnetoconductance
crossover. \newblock {\em Phys Rev B} 2013; \textbf{87}:241401.

\bibitem {Hess_prl_1990}Hess HF, Robinson RB, and Waszczak JV. \newblock
Vortex-core structure observed with a scanning tunneling microscope.
\newblock {\em Phys Rev Lett} 1990; \textbf{64}:2711--2714.

\bibitem {Gygi_prb_1990}Gygi F and Schluter M. \newblock Electronic tunneling
into an isolated vortex in a clean type-II superconductor.
\newblock {\em Phys Rev B} 1990; \textbf{41}:822--825.

\bibitem {Wu_arxiv_2016}Wu X, Liang Y and Fan H \textit{et al}. \newblock
Nematic orders and nematicity-driven topological phase transition in FeSe.
\newblock {\em arXiv:1603.02055 [cond-mat.supr-con]}, 2016.

\bibitem {Hao_prb_2017}Hao N, Zheng F and Zhang P \textit{et al}. \newblock
Topological crystalline antiferromagnetic state in tetragonal FeS.
\newblock {\em Phys Rev B} 2017; \textbf{96}:165102.

\bibitem {Hirschfeld_rpp_2011}Hirschfeld PJ, Korshunov MM and Mazin II.
\newblock Gap symmetry and structure of Fe-based superconductors.
\newblock {\em Rep Prog Phys} 2011; \textbf{74}:124508.

\bibitem {Chubukov_arxiv_2014}Chubukov AV and Hirschfeld PJ. \newblock
Fe-based superconductors: seven years later.
\newblock {\em arXiv preprint arXiv:1412.7104}, 2014.
\end{thebibliography}

\end{document}